\documentclass[aps,prx,a4paper,notitlepage,reprint,superscriptaddress]{revtex4-1}

\usepackage{amsmath,amssymb,amsfonts} 
\usepackage{mathrsfs} 

\usepackage{mathtools} 
\usepackage{color}
\usepackage{todonotes}
\usepackage{graphicx,float}
\usepackage[colorlinks,
linkcolor=red,
citecolor=blue,
urlcolor=red]{hyperref}

\usepackage{cleveref}

\renewcommand{\t}[1]{\mathrm{#1}}
\newcommand{\SiN}{Si$_3$N$_4\,$}

\newcommand{\figref}[1]{Fig.~\ref{#1}}
\renewcommand{\eqref}[1]{Eq.~\ref{#1}}

\begin{document}
\title{Generalized dissipation dilution in strained mechanical resonators}
	
\author{S. A. Fedorov}
\affiliation{Institute of Physics (IPHYS), {\'E}cole Polytechnique F{\'e}d{\'e}rale de Lausanne, 1015
		Lausanne, Switzerland}
	
\author{N. J. Engelsen}
\affiliation{Institute of Physics (IPHYS), {\'E}cole Polytechnique F{\'e}d{\'e}rale de Lausanne, 1015 Lausanne, Switzerland}

\author{A. H. Ghadimi}
\affiliation{Institute of Physics (IPHYS), {\'E}cole Polytechnique F{\'e}d{\'e}rale de Lausanne, 1015 Lausanne, Switzerland}
	
\author{M. J. Bereyhi}
\affiliation{Institute of Physics (IPHYS), {\'E}cole Polytechnique F{\'e}d{\'e}rale de Lausanne, 1015 Lausanne, Switzerland}
	
\author{R. Schilling}
\affiliation{Institute of Physics (IPHYS), {\'E}cole Polytechnique F{\'e}d{\'e}rale de Lausanne, 1015 Lausanne, Switzerland}
	
\author{D. J. Wilson}
\affiliation{IBM Research --- Zurich, Sa\"{u}merstrasse 4, 8803 R\"{u}schlikon, Switzerland}
	
\author{T. J. Kippenberg}
\email{tobias.kippenberg@epfl.ch}
\affiliation{Institute of Physics (IPHYS), {\'E}cole Polytechnique F{\'e}d{\'e}rale de Lausanne, 1015 Lausanne, Switzerland}
\email{tobias.kippenberg@epfl.ch}
	
\begin{abstract}
Mechanical resonators with high quality factors are of relevance in precision experiments, ranging from gravitational wave detection and force sensing to quantum optomechanics. Beams and membranes are well known to exhibit flexural modes with enhanced quality factors when subjected to tensile stress. The mechanism for this enhancement has been a subject of debate, but is typically attributed to elastic energy being ``diluted" by a lossless potential. Here we clarify the origin of the lossless potential to be the combination of tension and geometric nonlinearity of strain. We present a general theory of dissipation dilution that is applicable to arbitrary resonator geometries and discuss why this effect is particularly strong for flexural modes of nanomechanical structures with high aspect ratios. Applying the theory to a non-uniform doubly clamped beam, we show analytically how dissipation dilution can be enhanced by modifying the beam shape to implement ``soft clamping", thin clamping and geometric strain engineering, and derive the ultimate limit for dissipation dilution.
\end{abstract}

\date{\today}
\maketitle
	

\begin{figure}[t]
\centering
\includegraphics[width=\columnwidth]{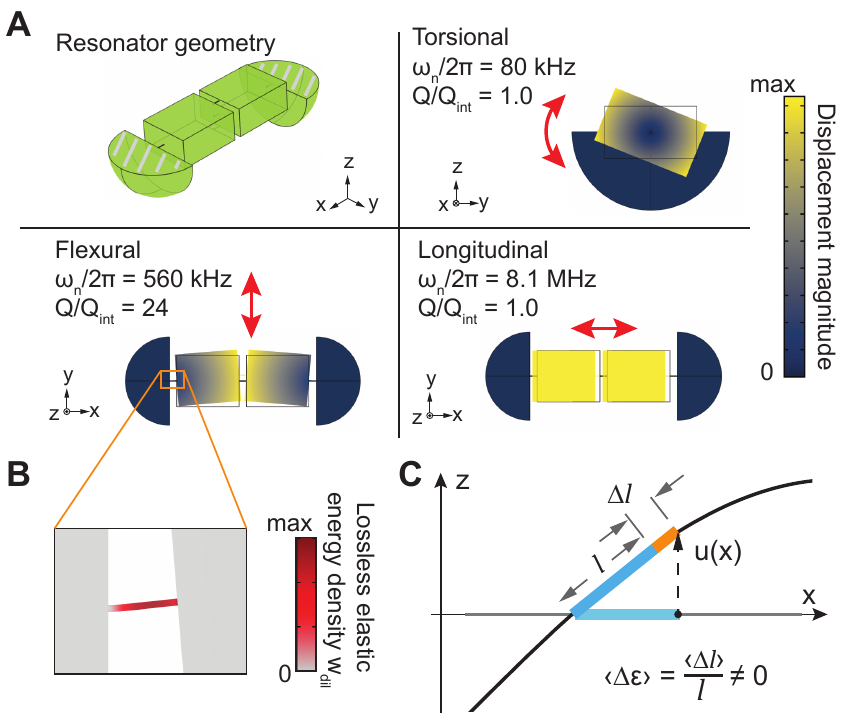}
\caption{A) Dissipation dilution factors for vibrational modes of a 3D resonator, doubly clamped to two quarter-sphere pads (hatched gray) and subjected to tension. The total length is 20 $\mu$m, the block size is $8.5\times7\times4$ $\mu$m, the bridge diameter is 100 nm and the material pre-strain is 0.4\%. B) Distribution of effectively lossless elastic energy in a thin bridge during flexural vibration. C) Schematic illustrating how the cycle-averaged dynamic strain $\langle \Delta \epsilon\rangle$ can be non-zero due to geometric nonlinearity.}
\label{fig:intro}
\end{figure}

\section{Introduction}
Mechanical resonators with high quality factors are of both fundamental and applied interest. They are employed in gravitational waves detector \cite{ju_detection_2000}, cavity optomechanics \cite{aspelmeyer_cavity_2014},
quantum \cite{andrews_bidirectional_2014} and classical \cite{bagci_optical_2014} signal conversion, tests of wavefunction collapse models \cite{vinante_improved_2017} and numerous sensing applications \cite{rugar_single_2004,yang_zeptogram-scale_2006}. In all these endeavors, dissipation can be a limiting factor. As known from the fluctuation-dissipation theorem \cite{callen_irreversibility_1951}, dissipation introduces noise, which limits force sensitivity, frequency stability and results in decoherence of quantum states.
Reduction of mechanical dissipation is practically challenging, however, because intrinsic and surface loss mechanisms are often not well understood or not possible to control. The quality factor, $Q$, of a mechanical resonator typically does not exceed the inverse of the material loss angle, $\phi$, characterizing the delay between stress and strain. Flexural modes of beams and membranes under tension are notable exceptions to this rule: they can have $Q$s far in excess of $1/\phi$ due to a phenomenon known as \emph{dissipation dilution}.\par

The origin of dissipation dilution has been a subject of debate. The concept was introduced in the gravitational wave community when, to explain the enhanced $Q$ of test mass suspension wire, Gonzalez \emph{et. al.}  \cite{gonzalez_suspensions_2000,gonzalez_brownian_1994} reasoned that the lossy elastic energy of the wire was ``diluted" by the conservative gravitational potential of the test mass. A decade later, similar behavior was observed in nanometric strings and membranes made of highly-strained materials (most notably, silicon nitride \cite{verbridge_high_2006,zwickl_high_2008,unterreithmeier_universal_2009}); however, the lack of an external potential in this case necessitated a rethinking of the physical model. In later works the quality factors of flexural modes of uniform beams \cite{Kotthaus_damping_PRL_2010} and membranes \cite{yu_control_2012} were calculated from a structural mechanics perspective and shown to be much greater than $1/\phi$---in excellent agreement with experiments \cite{Kotthaus_damping_PRL_2010,villanueva_surface_loss_2014,yu_control_2012,ghadimi_radiation_2017}. These results partially demystified dissipation dilution, but due to their lack of generality, the understanding of this effect remains incomplete. It is still not fully clear what causes dissipation dilution to emerge in a resonator (aside from the mere presence of tensile strain), if any modes except for flexural experience dilution and to what extent it can be engineered to produce practical high-$Q$ resonators.\par

Very recently, dissipation dilution has attracted significant interest as it enabled nanomechanical resonators, in the form of patterned membranes and beams, to achieve exceptionally high $Q$ factors \cite{tsaturyan_ultracoherent_2017,ghadimi_strain_2017}.
In particular, by localizing a beam mode away from its supports with a phononic crystal (the ``soft clamping" approach introduced by Tsaturyan et al. \cite{tsaturyan_ultracoherent_2017}) and using geometric strain engineering \cite{minamisawa_top-down_2012} to enhance strain in the beam constriction, $Q$ factors as high as $8\times 10^8$ were demonstrated at room temperature \cite{ghadimi_strain_2017}---surpassing even the highest values measured in macroscopic sapphire bars \cite{braginsky_systems_1985}.
These advances suggest that a more detailed understanding of dissipation dilution may be beneficial for optimizing existing designs and finding new ones, in addition to the open questions mentioned above.\par

Here we address these questions with a general and consistent theory which does not resort to the concept of an \emph{a priori} lossless potential. We derive the dissipation dilution factors for modes of a mechanical resonator of arbitrary geometry. We identify \emph{geometric nonlinearity} of strain in deformations to be a key component which, together with static strain, enables dissipation dilution. We extend the classic treatment of $Q$ dilution in flexural vibrations of a doubly-clamped beam to the case where the beam has a non-uniform width. Using this theory we show how a non-uniform width can be used to enhance $Q$ with three strategies: mode localization with phononic crystals \cite{tsaturyan_ultracoherent_2017}, both alone and in combination with adiabatic tapering \cite{ghadimi_strain_2017} and ``thin clamping", introduced here. We show that in a number of cases engineering dissipation dilution is related to geometric strain engineering \cite{zabel_top-down_2017,zhang_integrated_2015}. We also derive the ultimate limit of dissipation dilution set by the material yield strain. Our numerical analysis of beams is based on the one-dimensional Euler-Bernoulli equation and is in excellent agreement with a full 3D treatment. The numerical routines for nanobeam $Q$ factor calculations are implemented in a freely available \texttt{Mathematica} package \cite{mathematica_package}.\par

\section{Geometric origin of dissipation dilution}
Dissipation dilution is commonly illustrated by a harmonic oscillator subjected to an external lossless potential \cite{gonzalez_suspensions_2000}, as in the case of optically-trapped mirrors \cite{corbitt_optical_2007,ni_enhancement_2012} or massive pendula in a gravitational field \cite{gonzalez_suspensions_2000}. If $\omega_\t{int}$ is the oscillator natural frequency, $\phi$ is its loss angle \cite{saulson_thermal_1990} and $\omega_\t{dil}$ is the frequency of motion in the lossless potential, then the oscillator $Q$ factor is increased compared to the intrinsic value $Q_\t{int}\equiv1/\phi$ by the ``dilution factor",
\begin{equation}\label{eq:QdilToyModel}
D_Q\equiv\frac{Q}{Q_\t{int}}=\frac{\omega_\t{int}^2+\omega_{\t{dil}}^2}{\omega_\t{int}^2}.
\end{equation}
For flexural vibrations of tensioned beams or membranes, the $Q$ enhancement takes place similarly to \eqref{eq:QdilToyModel} with the important distinction that here the potential energy is stored only as elastic energy. Instead of introducing an external potential, the elastic energy is divided into lossy ``bending" and lossless ``tension" parts \cite{Kotthaus_damping_PRL_2010,yu_control_2012}, related to the curvature and gradient of the mode shape, respectively. It is not evident a priori, however, how to make this separation in a general case and under which conditions the lossless part of energy is non-zero. Here we answer both questions and show that the effectively lossless elastic energy emerges if two conditions are satisfied: a) static strain is non-zero in the resonator and b) the average of strain variation over the oscillation period is non-zero, i.e. the geometric nonlinearity of strain is significant.\par

We now derive the dissipation dilution factor of a generic vibrational mode. For this we compute the $Q$ factor as the ratio of the elastic energy stored by the mode to the energy dissipated per vibrational period. We assume that static deformation is present in the structure along with a part oscillating at the frequency $\omega_n$. Denoting the total deformation field as $\tilde{U}_i(x,y,z,t)$ ($i=x,\, y,\, z$), the strain tensor $\tilde{e}_{ij}$ \cite{landau_theory_1970} is given by
\begin{equation}\label{eq:strainTensor}
\tilde{e}_{ij}=\frac{1}{2}\left(\frac{\partial \tilde{U}_i}{\partial x_j}+\frac{\partial \tilde{U}_j}{\partial x_i}+\frac{\partial \tilde{U}_l}{\partial x_i}\frac{\partial \tilde{U}_l}{\partial x_j}\right),
\end{equation}
where summation over repeating indices is implied. The last term in \eqref{eq:strainTensor} is nonlinear in the displacement and can be identified as the geometric nonlinearity. We emphasize here that this nonlinearity is not due to a nonlinear stress-strain relation and is not always negligible even for infinitesimally small vibrations.\par

The strain tensor can be split into static $e_{ij}$ and time-dependent $\Delta e_{ij}(t)$ contributions
\begin{equation}
\tilde{e}_{ij}(t)=e_{ij}+\Delta e_{ij}(t).
\end{equation}
For brevity, when treating the 3D case we present a simplified model where Poisson's ratio, $\nu$, is neglected, so that the stress-strain relation is given by
\begin{equation}
\tilde{\sigma}_{ij}[\omega]=E e^{-i\phi}\tilde{e}_{ij}[\omega].
\end{equation}
A full treatment accounting for Poisson's ratio can be found in the Supplementary Information and $\nu$ is included below when treating flexural modes of beams.\par

We find the time-averaged elastic energy density stored by the mode as
\begin{multline}
\langle \Delta w_\t{el}(t)\rangle=E\frac{\langle\tilde{e}_{ij}(t)\tilde{e}_{ij}(t)\rangle}{2}-E\frac{e_{ij}e_{ij}}{2}=\\
E\left(e_{ij}\langle\Delta e_{ij}(t)\rangle+\frac{\langle\Delta e_{ij}(t)\Delta e_{ij}(t)\rangle}{2}\right),
\end{multline}
and the dissipated power density $p_{\t{diss}}$ as
\begin{equation}\label{eq:pDiss}
p_{\t{diss}}=\left\langle\tilde{\sigma}_{ij}(t)\,(\tilde{e}_{ij})'_t(t)\right\rangle=\omega_n\,\phi\, E\langle \Delta e_{ij}(t)\Delta e_{ij}(t)\rangle.
\end{equation}
The dilution factor of the vibrational mode is given by the ratio of the resonator quality factor to $Q_\t{int}$ as
\begin{equation}\label{eq:dilutionGeneral}
D_Q=1+\frac{\int 2 e_{ij}\langle\Delta e_{ij}(t)\rangle dV}{\int\langle\Delta e_{ij}(t)\Delta e_{ij}(t)\rangle dV}.
\end{equation}
\par

\eqref{eq:dilutionGeneral} reveals the peculiar effect of static strain $e_{ij}$ on dissipation. If the static strain is zero then $D_Q=Q/Q_\t{int}=1$ irrespective of the mode shape (we emphasize that corrections due to the imaginary part of Poisson’s ratio are here neglected). In contrast, $D_Q$ can be higher (or lower) than unity if $e_{ij}\neq 0$ and $\langle\Delta e_{ij}(t)\rangle\neq 0$, the latter being possible due to geometric nonlinearity in \eqref{eq:strainTensor}.\par

Comparing \eqref{eq:dilutionGeneral} to \eqref{eq:QdilToyModel}, one recognizes
\begin{equation}
\langle W_\t{dil}(t)\rangle\equiv E\int e_{ij}\langle\Delta e_{ij}(t)\rangle dV
\end{equation}
as an effectively lossless potential that generalizes the ``tension energy" in treatment of beams and membranes \cite{gonzalez_brownian_1994,yu_control_2012}. The lossy part of the energy is given by
\begin{equation}
\langle W_\t{lossy}(t)\rangle\equiv \frac{E}{2}\int \langle\Delta e_{ij}(t)\Delta e_{ij}(t)\rangle dV,
\end{equation}
which generalizes the ``bending energy" \cite{gonzalez_brownian_1994,yu_control_2012} and corresponds to $\omega_\t{int}^2$ in \eqref{eq:QdilToyModel}. Unlike the toy model, however, $W_\t{lossy}$ in general depends on the static strain, which implies that the intuitive picture that tension increases stored energy without affecting dissipation is not correct in general.\par

To give an example, we apply \eqref{eq:dilutionGeneral} to a doubly-clamped 3D resonator made of pre-strained material as shown in \figref{fig:intro}A and calculate dilution factors for a few representative modes from different families. It can be seen that among these modes only the flexural ones experiences dissipation dilution, whereas the torsional and longitudinal modes do not. A visualization of lossless energy density $\langle w_\t{dil}(t)\rangle$ in \figref{fig:intro}B shows that the lossless potential is concentrated in thin bridges between the blocks. This is explained by a) static strain concentration in constrictions and b) relatively large geometric nonlinearity of strain in flexural deformations, as opposed to torsional or longitudinal deformations.\par

Strong dissipation dilution of flexural modes in high-aspect-ratio beams and membranes \cite{zwickl_high_2008,Kotthaus_damping_PRL_2010} is thus due to the combination of tension and a large geometrically nonlinear contribution to the dynamic strain. The latter can be illustrated by considering flexural deformation of an idealized infinitely thin beam shown in \figref{fig:intro}C. If the beam is oriented along the $x$-axis and vibrates along the $z$-direction with magnitude $u$, only the diagonal component $\tilde{e}_{xx}\equiv \tilde{\epsilon}$ is relevant and the dynamic variation of strain is quadratic (i.e. fully nonlinear) in the displacement magnitude:
\begin{equation}\label{eq:strainVarBeam}
\Delta\epsilon(x,t)=(\tilde{u}'_x(x,t))^2/2.
\end{equation}
The role of geometric nonlinearity of strain in dissipation dilution provides a warning: it is not correct to assume that the mere presence of tensile strain in a mechanical resonator increases its $Q$---for example, torsional modes of the same structures that have high-$Q$ flexural modes usually do not experience any appreciable dissipation dilution (see \figref{fig:intro}A).\par

\section{Dissipation dilution of beam resonators}
For the rest of the paper we consider in detail the flexural modes of beams, as extreme dissipation dilution is achievable in this case and it is possible to obtain analytical results \cite{gonzalez_brownian_1994,villanueva_surface_loss_2014}. Applying \eqref{eq:dilutionGeneral} we arrive at a dilution factor given by
\begin{equation}\label{eq:dilutionBeam}
D_Q=1+\frac{\int 2 \epsilon\langle\Delta \epsilon(t)\rangle dV}{\int\langle\Delta \epsilon(t)^2\rangle dV},
\end{equation}
where $\epsilon$ is the static strain along the beam, terms proportional to $\epsilon\langle\Delta \epsilon(t)\rangle$ and $\langle\Delta \epsilon(t)^2\rangle$ correspond to the lossless ``tension" and lossy ``bending" energy, respectively \cite{gonzalez_brownian_1994,yu_control_2012}---both are of elastic origin. Note that while \eqref{eq:dilutionGeneral} neglects Poisson's ratio, \eqref{eq:dilutionBeam} does not, and is formally exact in the 1D case.\par

So far we have not made any assumptions about the beam cross-section, but in the following we focus on geometries directly accessible by nanofabrication. Specifically, we assume that the beams are made of a suspended film with thickness $h$ and pre-strain $e_{xx}=e_{yy}=\epsilon_\t{film}$ (which redistributes upon suspension). The beam width $w(x)$ is in general non-uniform and its variation can be used to improve vibrational quality factors.\par

For modes of a uniform rectangular beam evaluation of \eqref{eq:dilutionBeam} yields the well-known result \cite{gonzalez_brownian_1994,villanueva_surface_loss_2014}
\begin{equation}\label{eq:DQrectBeam}
D_{Q,n}=\frac{1}{2\lambda+\pi^2 n^2\lambda^2}.
\end{equation}
Here $n$ is mode number and $\lambda$ is defined as \cite{yu_control_2012,villanueva_surface_loss_2014}
\begin{equation}\label{eq:lambdaDef}
\lambda^2=\frac{1}{12\epsilon_{\t{avg}}}\frac{h^2}{l^2},
\end{equation}
where $\epsilon_{\t{avg}}$ is the volume-averaged static tensile strain and $l$ is the beam length.\par

\begin{figure}[t]
	\centering
	\includegraphics[width=\columnwidth]{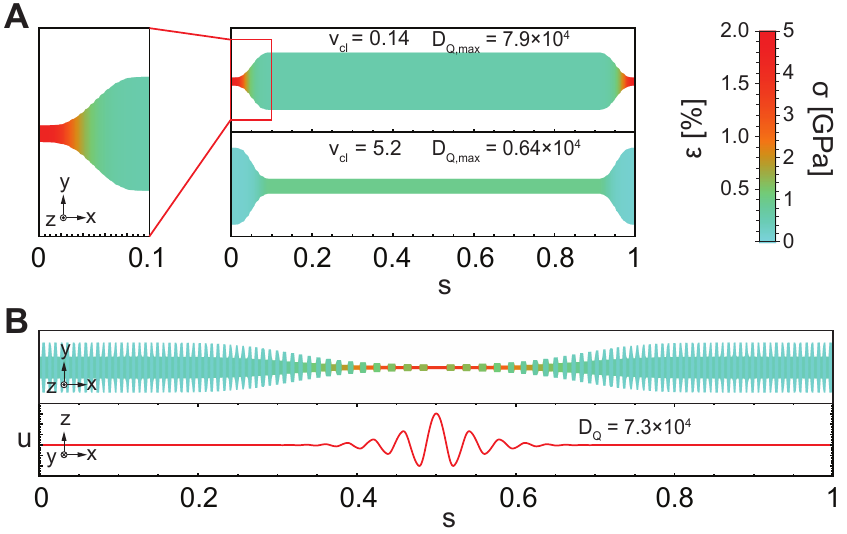}
	\caption{Geometry, strain distribution and $D_Q$ in micropatterned beams, illustrating the concepts of soft-clamping, thin clamping and strain-engineering.
	Dilution factors ($D_Q$) are calculated assuming beam length $l=3$ mm and thickness $h=20$ nm.  A) Beams with thin (above) and thick (below) clamps, resulting in enhanced and reduced dissipation dilution, respectively. $D_{Q,\t{max}}$ is maximum over modes. B) Strain (top) and localized mode displacement field (bottom) in a tapered phononic crystal beam.}
	\label{fig:thinClamp}
\end{figure}

The derivation of \eqref{eq:DQrectBeam} is based on a key insight: the flexural modes of a beam contain two vastly different length scales \cite{gonzalez_suspensions_2000,gonzalez_brownian_1994}. Away from the clamping points (clamps), modes form standing waves with wavelengths on the order of $2l/n$, while near the clamping points they experience sharp bending at the length scale of $\lambda l$, which is responsible for fulfilling the clamped boundary conditions $u'=0$. As a result, the majority of the elastic energy is distributed over the mode away from the clamping points, while the small regions around them make a large (dominant for lowest-frequency modes) contribution to the intrinsic losses \cite{yu_control_2012,Kotthaus_damping_PRL_2010}. The energy dissipation around the clamping points is commonly referred to as ``clamping losses" \cite{yu_control_2012}, which, should not be confused with losses due to modal coupling to the supporting frame \cite{wilson-rae_high-q_2011,cole_phonon-tunnelling_2011,tsaturyan_tunneling_loss_2014} or acoustic radiation \cite{chan_optimized_2012,ghadimi_radiation_2017}. In the following we refer to the intrinsic loss occurring away from the clamps as ``distributed contribution".\par

We now generalize the multi-length scale approach for the case of non-uniform beams and derive dissipation dilution factors as (see details in SI)
\begin{equation}\label{eq:DQgeneral}
	D_{Q,n}=\frac{1}{2\alpha_n\lambda+\beta_n\Omega^2_n\lambda^2},
\end{equation}
where we introduced dimensionless frequency of $n$-th mode $\Omega_n$ given by
\begin{equation}\label{eq:dimlessFreqDef}
\Omega_n^2=\frac{\rho l^2 \omega_n^2}{\epsilon_{\t{avg}}E},
\end{equation}
and beam shape-dependent clamping and distributed loss coefficients $\alpha_n$ and $\beta_n$ are found as
\begin{align}
&\alpha_n=\frac{\sqrt{v_{\mathrm{cl}}}(u_{\t{cl},n}')^2}{2\Omega^2_n\left(\int_0^1v(s)u_n(s)^2ds\right)},\label{eq:alphaDef}\\
&\beta_n=\frac{\int_{0}^{1}v(s)^3 u_n(s)^2ds}{\int_0^1v(s)u_n(s)^2ds}.\label{eq:betaDef}
\end{align}
Here $s=x/l$ is the scaled coordinate along the beam, $u_n(s)$ is the mode shape, $v(s)=w(s)/w_\t{avg}$ is the beam width variation normalized to its average width and quantities with subscript ``cl" are computed near the clamps (see SI).\par

Dissipation dilution of a non-uniform beam can be discussed entirely in terms of the reduction of the $\alpha_n$ and $\beta_n$ coefficients by varying the beam shape $w(x)$; however, some results are more intuitively interpreted from the prospective of geometric strain engineering \cite{zabel_top-down_2017,ghadimi_strain_2017,zhang_integrated_2015}, a technique that exploits relaxation of a suspended film to locally enhance the strain. Formally, the treatment in terms of the transverse beam shape, $w(x)$, or the static strain distribution along the beam, $\epsilon(x)$, is equivalent as these quantities are uniquely related as (see SI for details)
\begin{equation}\label{eq:strainDistr}
\epsilon(x)/\epsilon_{\t{avg}}=w_\t{avg}/w(x),
\end{equation}
through the condition that the tension force must be constant along the beam.

\section{Dissipation dilution limit}
Before showing how dissipation dilution can be enhanced in a non-uniform beam, we derive a rigorous upper bound for $D_Q$. This bound is set by the yield strain, material parameters, beam thickness and the frequency of vibration, but does not depend on the beam length nor the mode order. We assume that the clamping losses are negligible ($\alpha_n=0$) and evaluate the distributed loss coefficient $\beta_n$ using the strain-width relation (\eqref{eq:strainDistr}) and the condition that the maximum strain in the beam cannot exceed the yield strain $\epsilon_\t{yield}$. As a result we obtain (see SI for details)
\begin{equation}
\beta_n\ge \left(\frac{\epsilon_\t{avg}}{\epsilon_\t{yield}}\right)^2,
\end{equation}
and thus the ultimate dissipation dilution bound is given by
\begin{equation}\label{eq:DQmax}
D_Q\le \frac{12E\epsilon_\t{yield}^2}{\rho h^2\omega^2}.
\end{equation}
This limit is formally equivalent to the dissipation dilution of a clampless uniform beam strained to the yield strain.

\section{Non-uniform beams with enhanced dissipation dilution}

\begin{figure}[t]
	\centering
	\includegraphics[width=\columnwidth]{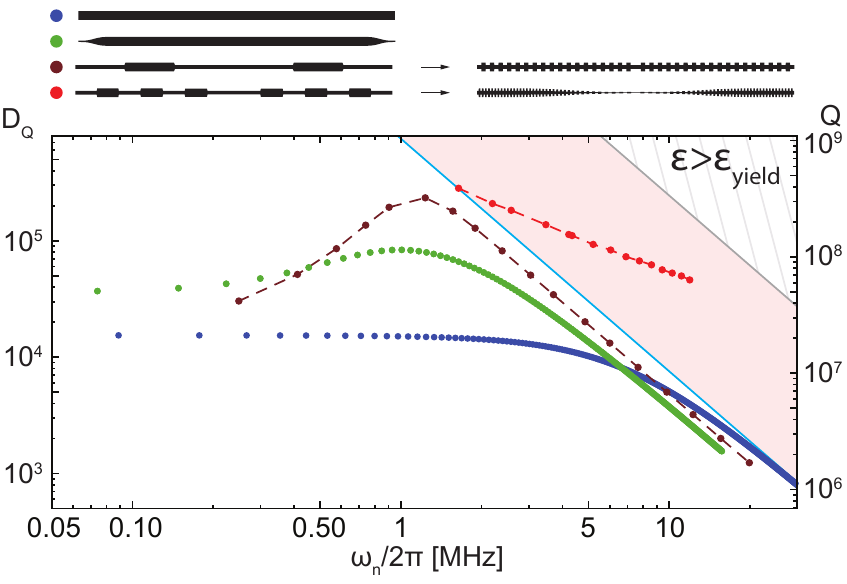}
	\caption{Dissipation dilution in beams with different transverse profiles, assuming a fixed length $l=3$ mm and thickness $h=20$ nm. Points correspond to $D_Q$ (left axis) and $Q$ (right axis) for specific flexural modes, assuming $Q_\t{int} = 1.4\times 10^3$.  Blue and green points correspond to modes of uniform and thin-clamped ($v_\t{cl}=0.14$) beams. Dark red and red points correspond to localized modes of PnC beams and tapered PnC beams, respectively. Note that each localized mode corresponds to a different beam profile. Blue line: ideal limit for a soft-clamped beam (\eqref{eq:DQclampless}). Gray line: ideal limit for a clamp-free beam strained to the yield point (\eqref{eq:DQmax}).}
	\label{fig:methodsCmp}
\end{figure}

We consider three beam designs that produce vibrational modes with enhanced dissipation dilution compared to uniform beam --- phononic crystal (PnC) beams, beams with thin clamps and tapered PnC beams. We first analytically estimate the attainable $D_Q$s with these designs and then numerically calculate them by solving the Euler-Bernoulli equation \cite{mathematica_package} (see SI). Numerical calculations are presented in \figref{fig:methodsCmp} for beams with length $l=3$ mm and thickness $h=20$ nm. We show dissipation dilution factors, which are material independent, along with absolute $Q$ factors assuming parameters typical to stoichiometric \SiN films ($E=250$ GPa, $\nu=0.23$, $\sigma_\t{film}=1.14$ GPa, $Q_\t{int}=1.4\times10^3$ for $h=20$ nm), a well-established material for strained nanomechanics \cite{villanueva_surface_loss_2014}. Note that with these extreme parameters the maximum dilution factor is large ($D_Q>10^4$) even for a uniform beam.\par

The first strategy we consider is soft clamping \cite{tsaturyan_ultracoherent_2017,ghadimi_strain_2017} --- suppression of clamping losses by localizing a flexural mode in a phononic crystal. A 1D phononic crystal can be formed by periodically modulating the beam width \cite{ghadimi_radiation_2017} (with $w_\t{max}=2w_\t{min}$ for the design in \figref{fig:methodsCmp}). Localized modes of a PnC beam can closely approach the performance of idealized clamp-less beams, with dilution factors given by
\begin{equation}\label{eq:DQclampless}
D_Q=\frac{12E\epsilon_{\t{film}}^2}{(1-\nu)^2\rho h^2\omega^2}.
\end{equation}
Here Poisson's ratio accounts for relaxation of film stress in transverse direction upon suspension. Importantly, the strong suppression of mechanical mode amplitude near the clamps requires a large number of PnC unit cells and thus a high order $n$ of the localized mode. For high-order modes, distributed losses increase as $n^2$ due to increased bending curvature for shorter acoustic wavelengths and at some point exceed the suppressed clamping losses. These trends can be seen in \figref{fig:methodsCmp}, where the $D_Q$ factor of the localized mode is plotted versus frequency. $D_Q$ can be optimized by changing the localized mode order $n$ while keeping all the parameters except for the unit cell length fixed. The amplitude of a localized mode decays exponentially with the distance from the defect, such that the clamping loss coefficient can be estimated as $\alpha_n=e^{-(n-1)/n_L}$,
where $n_L$ is the mode amplitude decay length in units of acoustic half-wavelengths. Optimization of $D_Q$ in \eqref{eq:DQgeneral} with respect to $n$, yields
\begin{equation}\label{eq:DQmaxSoftClamp}
	D_{Q,\mathrm{max}}\approx\frac{1}{\pi^2n_{\mathrm{max}}^2\lambda^2},
\end{equation}
where $n_{\mathrm{max}}$ is the optimum localized mode order that increases only logarithmically slowly with $1/\lambda$ (see SI for the explicit expression). This result demonstrates that patterning a beam with a phononic crystal can provide an improvement in $D_Q$ by a factor of $\sim 1/(n_\t{max}^2\lambda)$ compared to a uniform beam of the same size. Note that the maximum attainable $D_Q$ is far below $1/\lambda^2$---the enhancement expected from clamping loss suppression for a fundamental mode---as $n_\t{max}$ is in practice much greater than 1. It also follows from \eqref{eq:DQmaxSoftClamp} that in order for soft clamping to provide an increased quality factor, $\lambda$ needs to be much smaller than 1, i.e. dissipation dilution factors needs to be high even for non-localized modes.\par

The second strategy we consider is reduction of the beam width near the clamps, $v_\t{cl}=w(0)/w_\t{avg}$, in order to create local strain enhancement in clamping regions (see \figref{fig:thinClamp}A top). \eqref{eq:alphaDef} shows that $\alpha_n$ is proportional to $\sqrt{v_\t{cl}}$ and thus can be reduced by thinning down the clamps ($u'_{\t{cl},n}$ and $\Omega_n$ are almost unaffected by $v_\t{cl}$ as long as the clamping region length is small). This can be interpreted as an effective decrease of $\lambda$ over the clamping region to
\begin{equation}
	\lambda_\t{cl}=\sqrt{h^2/12\epsilon_\t{cl} l^2},
\end{equation}
where $\epsilon_\t{cl}=\epsilon_\t{avg}/v_\t{cl}$ is the local strain. The dissipation dilution of beams with thin clamps is thus given by
{\begin{equation}\label{eq:DQThinClamp}
D_{Q,n}\approx \frac{1}{2\lambda_\t{cl}+(n\pi)^2\lambda^2}.
\end{equation}
In contrast to the PnC approach, thin-clamping beams are predicted to have improved quality factors for low-order beam modes, including the fundamental mode (see \figref{fig:methodsCmp}, green points).\par

One caveat needs to be mentioned when considering the effect of local strain on dissipation dilution: geometric concentration of strain in one region unavoidably results in the reduction of strain elsewhere. To improve dilution factors beyond those of a uniform beam, the region(s) of enhanced strain must overlap with the region(s) which dominate dissipation in the vibrational mode, in this case the clamps. A common beam geometry which does not satisfy this requirement, a beam with filleted (thick) clamping points, is shown in the bottom of \figref{fig:thinClamp}A. This result is at odds with recently reported enhanced $Q$s in trampoline membranes with filleted tethers \cite{norte_mechanical_2016}.\par

In both uniform PnC and thin-clamped beams, the clamping loss is reduced, but distributed loss is not. The latter can be addressed by co-localization of both flexural mode and strain away from the clamps as shown in \figref{fig:thinClamp}B. Following the strategy described in \cite{ghadimi_strain_2017}, here the width of the PnC is changed cell-wise according to
\begin{equation}
w_{\t{cell,}i}\propto 1-(1-a)\exp(-i^2/i_0^2),
\end{equation}
where $i=0,\,1\,...$ is the cell index starting from the beam center, $a$ and $i_0$ respectively define the transverse and longitudinal sizes of the waist region. Importantly, the PnC cell lengths must also be scaled proportional to $1/\sqrt{w_\t{cell}}$ in order to compensate for the bandgap frequency shift due to the non-uniform strain distribution.\par

An estimate of $D_Q$ for the tapered PnC is obtained by assuming that the mode is localized in the waist region of width $v_\t{waist}$ and that clamping losses are negligible:
\begin{equation}\label{eq:DQscse}
	D_{Q,n}\approx\frac{1}{\Omega^2_\t{waist}\lambda^2_\t{waist}},
\end{equation}
where
\begin{align}
	&\Omega_\t{waist}=\sqrt{\rho l^2 \omega^2/(\epsilon_\t{waist}E)},\\
	&\lambda_\t{waist}=\sqrt{h^2/(12\epsilon_\t{waist}l^2)},
\end{align}
and $\epsilon_\t{waist}=\epsilon_\t{avg}/v_\t{waist}$. It follows that by increasing the waist strain to yield value, the ultimate limit of dissipation dilution (\eqref{eq:DQmax}) is attainable with tapered PnC beam designs, in contrast to the previous two methods. \par
	
A practical limitation for dissipation dilution enhancement by strain concentration in this case originates from the tradeoff between $\epsilon_\t{waist}$ and the waist length. Substantially increased strain is only achievable over a small fraction of the beam length, therefore only short-wavelength and high-frequency modes can benefit from such global geometric strain engineering. In \figref{fig:methodsCmp} we plot $D_Q$ versus frequency for localized modes of tapered beams, where the taper waist has been adjusted to match the wavelength of the localized mode. It can be seen that as the mode frequency increases, its dilution is progressively enhanced relative to conventional soft-clamped modes (red points).

\section{Conclusions and outlook}
We have presented a theoretical framework to analyze the quality factors of strained mechanical resonators of arbitrary three dimensional geometry and shown that a lossless contribution to the elastic energy ---giving rise to $Q$-enhancement by dissipation dilution --- emerges in the presence of static strain and geometric nonlinearity. High aspect ratio beams and membranes can produce particularly large dissipation dilution, though it is not impossible that other geometries can do it as well.\par

For the specific case of variable cross-section beams subjected to axial tension we presented an analytical model. We showed that by corrugating the beam it is possible to create modes with quality factors enhanced by more than an order of magnitude compared to a uniform beam. We interpret the $Q$ enhancement in terms of clamping loss suppression and local strain engineering, deriving the limits of each approach, and estimating practically achievable absolute $Q$ factors for beams made of high-stress \SiN. The numerical results reported for beams were obtained using a freely available \texttt{Mathematica} package \cite{mathematica_package}. \par

We note that while \SiN is currently the most popular material for strained nanomechanics --- particularly for applications in optomechanics \cite{thompson_strong_2008,wilson_cavity_2009,purdy_observation_2013,wilson_measurement-based_2015} --- the principles described here apply to resonators made of any material under strain, whether produced by external force \cite{verbridge_macroscopic_2007}, lattice mismatch (e.g. during epitaxial growth) \cite{cole2014tensile} or mismatch of thermal expansion coefficients \cite{regal_measuring_2008}.

\section*{Acknowledgements}
We thank Alexander Tagantsev for useful discussions. This work was supported by the EU Horizon 2020 Research and Innovation Program under grant agreement no. 732894 (FET Proactive HOT), the SNF Cavity
Quantum Optomechanics project (grant no. 163387) and DARPA grant HR0011181003. M.J.B. is supported by MSCA ETN-OMT
(grant no. 722923). T.J.K acknowledges support from ERC AdG (QuREM, grant no. 320966). Code to reproduce data in \figref{fig:thinClamp} and \figref{fig:methodsCmp} is available on Zenodo\cite{mathematica_package}.

%

\clearpage

\onecolumngrid
\setcounter{equation}{0}
\setcounter{figure}{0}
\setcounter{section}{0}

\renewcommand{\thefigure}{S\arabic{figure}}
\renewcommand{\theequation}{S\arabic{equation}}

\begin{center}
	\Large \textbf{Supplementary information for ``Generalized dissipation dilution in strained mechanical resonators"}
\end{center}

\section{Dissipation dilution in a generic mechanical resonator made of isotropic anelastic material}
In the main manuscript Poisson's ratio is neglected to derive a simplified expression derived for the dissipation dilution of modes of a 3D resonator. Here we remove this simplification and present a more general formula taking Poisson's ratio, $\nu$, into account. As in the main text, $E$ is the Young's modulus, $\phi$ is the loss angle and summation over repeating indices is assumed.\par

The time-dependent displacement field is denoted by $\tilde{U}_i(x,y,z,t)$, where $i=x,\,y,\,z$ is the coordinate index. The train field is derived from it as\cite{landau_theory_1970_SI}
\begin{equation}
\tilde{e}_{ij}=\frac{1}{2}\left(\frac{\partial \tilde{U}_i}{\partial x_j}+\frac{\partial \tilde{U}_j}{\partial x_i}+\frac{\partial \tilde{U}_l}{\partial x_i}\frac{\partial \tilde{U}_l}{\partial x_j}\right),
\end{equation}
and the stress $\tilde{\sigma}_{ij}$ is related to strain via Hooke's law\cite{landau_theory_1970_SI}
\begin{equation}
\tilde{\sigma}_{ij}=\frac{E}{1+\nu}\left(\tilde{e}_{ij}+\frac{\nu}{1-2\nu}\tilde{e}_{ll}\delta_{ij}\right),
\end{equation}
where $\delta_{ij}$ is the Kronecker delta. Following the main text, we now assume that the deformation field consists of a static part, $U_i(x,y,z)$, and a dynamic part due to mechanical vibrations, $\Delta U_{i,n}(x,y,z,t)$, which is given by
\begin{equation}
\Delta U_i(x,y,z,t) = \Delta U_{i,n}(x,y,z)e^{-i\omega_n t}+c.c,
\end{equation}
where $\Delta U_{i,n}(x,y,z)$ and $\omega_n$ is the complex envelope and frequency of $n$-th mode. Strain, stress and elastic energy can be separated into static and time-dependent contributions accordingly:
\begin{align}
&\tilde{e}_{ij}(t)=e_{ij}+\Delta e_{ij}(t),\\
&\tilde{\sigma}_{ij}(t)=\sigma_{ij}+\Delta \sigma_{ij}(t),\\
&\tilde{w}(t)=w+\Delta w(t).
\end{align}

\noindent The instantaneous elastic energy density is then given by
\begin{equation}
\tilde{w}=\frac{1}{2}\tilde{\sigma}_{ij}\tilde{e}_{ij}=\frac{E}{2(1+\nu)}\left(\tilde{e}_{ij}\tilde{e}_{ij}+\frac{\nu}{1-2\nu}(\tilde{e}_{ll})^2\right),
\end{equation}
and the average of its variation, $\Delta w(t)$, which is the elastic energy stored by the vibrational mode, is found as
\begin{equation}\label{eq:averEnVarSI}
\begin{split}
\langle \Delta w(t)\rangle=&\frac{1}{2}(\sigma_{ij}\langle \Delta e_{ij}(t)\rangle+e_{ij}\langle\Delta\sigma_{ij}(t)\rangle+\langle\Delta\sigma_{ij}(t)\Delta e_{ij}(t)\rangle) \\
=&\frac{E}{2(1+\nu)}\left((2e_{ij}\langle\Delta e_{ij}(t)\rangle+\langle\Delta e_{ij}(t)\Delta e_{ij}(t)\rangle)+\frac{\nu}{1-2\nu}(2e_{ll}\langle\Delta e_{kk}(t)\rangle+\langle(\Delta e_{kk}(t))^2\rangle)\right).
\end{split}
\end{equation}

\noindent We can then find the dissipated power density as
\begin{equation}
p_\t{diss}=\left\langle\tilde{\sigma}_{ij}\frac{\partial\tilde{e}_{ij}}{\partial t}\right\rangle=\sigma_{ij}\left \langle\frac{\partial}{\partial t}\Delta e_{ij}(t)\right\rangle+\left\langle\Delta\sigma_{ij}(t)\frac{\partial }{\partial t}\Delta e_{ij}(t)\right\rangle.
\end{equation}
Here, the second term, $\langle\Delta\sigma_{ij}(t)\partial\Delta e_{ij}(t)/\partial t\rangle$, yields non-zero dissipated power if a delayed strain response to stress is introduced as a perturbation by the substitution $\Delta e_{ij}[\omega]\to (1+i\phi)\Delta e_{ij}[\omega]$ and the average over time is found using the unperturbed $\Delta e_{ij}$. Unlike \eqref{eq:averEnVarSI} for the stored energy, the extra term which arises in the presence of static deformation, $\sigma_{ij}\langle\partial\Delta e_{ij}(t)/\partial t\rangle$, is always zero as
\begin{equation}
\left \langle\frac{\partial}{\partial t}\Delta e_{ij}(t)\right\rangle=\frac{1}{T}\int_{0}^{T}\frac{\partial}{\partial t}\Delta e_{ij}(t)dt=\Delta e_{ij}(T)-\Delta e_{ij}(0)=0,
\end{equation}
where $T$ is the oscillation period. Overall the dissipated power density is found as
\begin{equation}\label{eq:pDissGeneralSI}
p_\t{diss}=\phi\omega_n \langle\Delta\sigma_{ij}(t)\Delta e_{ij}(t)\rangle=\phi \omega_n\frac{E}{(1+\nu)}\left(\langle\Delta e_{ij}(t)\Delta e_{ij}(t)\rangle+\frac{\nu}{1-2\nu}\langle(\Delta e_{kk}(t))^2\rangle\right).
\end{equation}
We find the quality factor of the mode from the stored energy and dissipation rate as
\begin{equation}
Q=\frac{2\omega_n\int \langle \Delta w(t)\rangle dV}{\int p_\t{diss} dV},
\end{equation}
and then find the dissipation dilution ratio as
\begin{equation}\label{eq:QIntDilSI}
D_{Q}=\frac{Q}{Q_\t{int}}=1+\frac{2\int (\sigma_{ij}\langle \Delta e_{ij}(t)\rangle+e_{ij}\langle\Delta\sigma_{ij}(t)\rangle) dV}{\int\langle \Delta\sigma_{ij}(t)\Delta e_{ij}(t)\rangle dV}=1+\frac{\langle W_\t{dil}(t)\rangle}{\langle W_\t{lossy}(t)\rangle},
\end{equation}
where $Q_\t{int}=1/\phi$ and the dilution and lossy energies are given, respectively, by\par

\begin{align}
&\langle W_\t{dil}(t)\rangle=\int (\sigma_{ij}\langle \Delta e_{ij}(t)\rangle+e_{ij}\langle\Delta\sigma_{ij}(t)\rangle) dV,\label{eq:WdilGeneralSI}\\
&\langle W_\t{lossy}(t)\rangle=\frac{1}{2}\int\langle \Delta\sigma_{ij}(t)\Delta e_{ij}(t)\rangle dV.\label{eq:WlossyGeneralSI}
\end{align}
\eqref{eq:WdilGeneralSI}-\ref{eq:WlossyGeneralSI} generalize the expressions for dilution and lossy elastic energies  presented in the main text for the case of non-zero Poisson's ratio and reproduce them if $\nu=0$.

\section{Derivation of dissipation dilution in a doubly clamped non-uniform beam}

\begin{figure}[t]
	\centering
	\includegraphics[width=\textwidth]{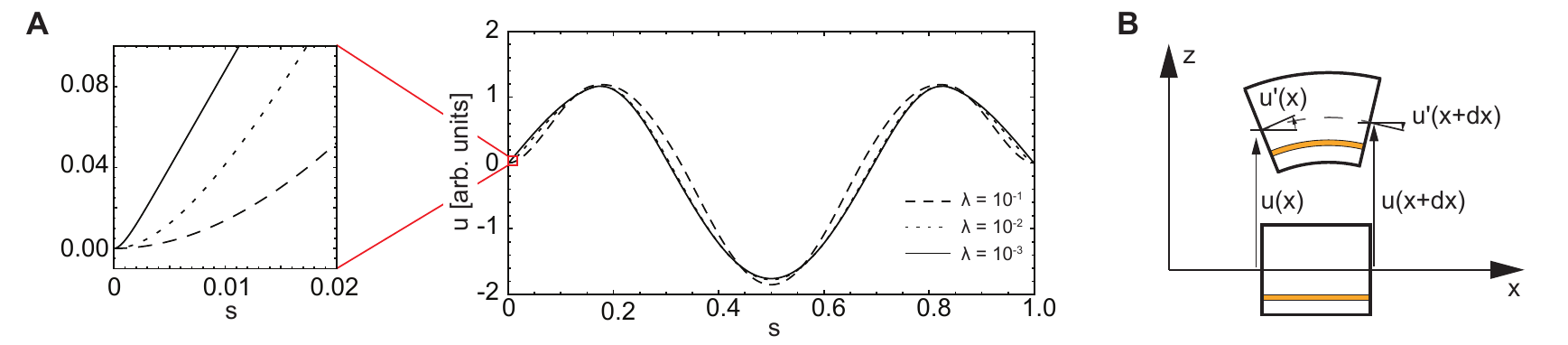}
	\caption{A) Shapes of third order flexural modes of beams assuming the values of $\lambda=10^{-1},\,10^{-3},\,10^{-3}$. The zoomed-in region shows the mode in the clamping regions, which illustrates an increase in mode curvature around clamps with the reduction of $\lambda$. B) Deformation of a segment vibrating beam.}
	\label{fig:bendIllSI}
\end{figure}

We consider flexural vibrations of thin doubly-clamped beam resonators and use the general result from the previous section to derive more useful expressions for quality factors and dissipation dilution factors. The beam will have dimensions $h,\, w$ and $l$ corresponding to the thickness ($z$-direction), width ($y$-direction) and length ($x$-direction). We assume that $h,\, w\ll l$, but we do not impose restrictions on the beam cross-section and do not assume $h$ and $  w$ are constant. The beam is suspended between two clamps and experiences a tensile force $T$, which creates an equilibrium axial strain of $\epsilon(x)\equiv e_{xx}$ and a stress given by $\sigma(x)\equiv\sigma_{xx}=E\, \epsilon(x)$. Due to the high aspect ratio of the beam, we can neglect stresses in all directions other than the $x$ axis.\par

We now consider the displacement of a beam segment in $z$ direction $u(x)\equiv\Delta U_z$, as illustrated in \figref{fig:bendIllSI}B. We then find the instantaneous variation of strain $\Delta\epsilon\equiv \Delta e_{xx}$ and stress $\Delta\sigma\equiv \Delta \sigma_{xx}$ to be given by
\begin{align}
	&\Delta\epsilon(x,y,z,t)=-u''_{xx}(x,t)z+\frac{(u'_x(x,t))^2}{2}\label{eq:epsilonSI},\\
	&\Delta\sigma(x,y,z,t) = E\, \Delta\epsilon(x,y,z,t).
\end{align}
The elastic energy density stored by the flexural mode is found as a sum of two terms
\begin{equation}\label{eq:WelSI}
	\langle \Delta w(t)\rangle=\frac{E}{2}(2\epsilon\langle \Delta\epsilon(t)\rangle+\langle\Delta\epsilon(t)^2\rangle)=\langle w_{\t{tens}}(t)\rangle+\langle w_{\t{bend}}(t)\rangle,
\end{equation}
where the first term is tension energy and the second is bending energy. Inserting \eqref{eq:epsilonSI} into \eqref{eq:WelSI} and performing integration over the beam volume, we find the total energy contributions
\begin{align}
	&\langle W_{\t{tens}}(t)\rangle=\int_{l}\frac{E}{2}A(x)\epsilon(x) (u'_x(x,t))^2dx
		=\frac{T}{2}\int_{l} (u'_x(x,t))^2dx,\label{eq:WelTensSI}\\
	&\langle W_{\t{bend}}(t)\rangle=\int_{l}\frac{E}{2}I(x)(u''_{xx}(x,t))^2dx.\label{eq:WelBendSI}
\end{align}
Here $I(x)=w(x)h(x)^3/12$ is the geometrical moment of inertia, $A(x)=w(x)h(x)$ is the cross-section area and we used the fact that the tension $T=EA(x)\epsilon(x)$ is constant along the beam. Provided that, according to \eqref{eq:pDissGeneralSI}, the dissipation power density is given by
\begin{equation}
	p_\t{diss}=\phi\,\omega_n E\langle\Delta\epsilon(t)^2\rangle=2\phi\,\omega_n\langle w_{\t{bend}}(t)\rangle.
\end{equation}
we find dissipation dilution factor of a flexural beam mode as
\begin{equation}
D_Q=1+\frac{\langle W_{\t{tens}}(t)\rangle}{\langle W_{\t{bend}}(t)\rangle}.
\end{equation}

\subsection{Nanobeams and equilibrium strain distribution in a suspended film}\label{sec:equlStrainSI}

\begin{figure}[t]
	\centering
	\includegraphics[width=0.5\textwidth]{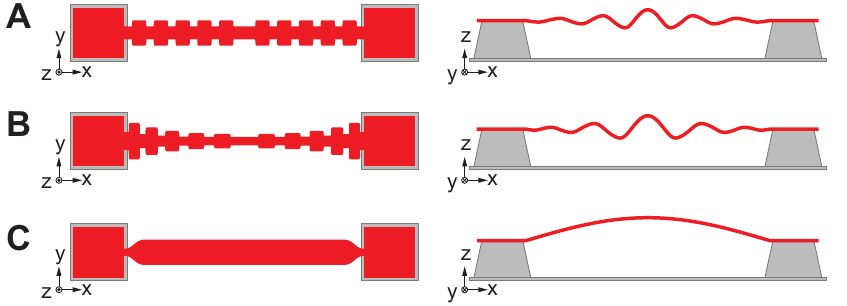}
	\caption{Beam resonator shapes with uniform thickness and non-uniform width demonstrating enhanced dissipation dilution. A) Phononic crystal beam with ``soft-clamped" localized mode, B) Tapered phononic crystal beam with soft-clamped mode and engineered local strain enhancement C) Beam with thin clamps and its fundamental mode.}
	\label{fig:beamsGeometrySI}
\end{figure}

Until now we have considered beams of arbitrary variable transverse cross-section. In the following we impose geometrical constraints, consistent with nanomechanical resonators fabricated by locally suspending a micropatterned thin film. Although qualitatively most of our conclusions are not affected by this assumption, it considerably simplifies notations while allowing the theory to be directly applied to a very broad range of practical high-strain resonators. In particular, we assume the $yz$ cross section of the beam to be rectangular, the width $w(x)$ be, in general, $x$-dependent and the thickness to be constant (representative geometries are shown in \figref{fig:beamsGeometrySI}). Strain can be present in a material film used for microresonator fabrication due to lattice mismatch\cite{cole2014tensileSI} between the film and substrate or by mismatch in their thermal expansion coefficients\cite{moridi_residual_2013_SI}. Upon suspension, the originally homogeneous strain inside the film is redistributed. The strain is locally enhanced in constrictions and reduced elsewhere\cite{minamisawa_top-down_2012_SI,zabel_top-down_2017_SI,capelle_polarimetric_2017_SI}.\par

The analysis of the vibrational properties of a beam in this case requires the axial tension force $T$ to be found first from the unsuspended film strain $\epsilon_{\t{film}}$. This can be done by noting that (a) the total elongation of the beam $\int_{0}^{l}\epsilon(x)dx$ is constant over the relaxation process, as it is defined by separation of the beam clamping points and (b) that the balance of tensile force requires
\begin{equation}\label{eq:tensForceBalSI}
\epsilon(x)w(x)=\t{const}=\frac{T}{hE}.
\end{equation}
\noindent From the initial condition
\begin{equation}
\int_{0}^{l}\epsilon(x)dx=\epsilon_{\t{film}}(1-\nu)l,
\end{equation}
\noindent where $\nu$ is the Poisson's ratio and the factor $(1-\nu)$ accounts for transverse relaxation of the strain, one finds the equilibrium tension as
\begin{equation}
T=\epsilon_{\t{film}}E(1-\nu)h\left(\frac{1}{l}\int_{0}^{l}\frac{1}{w(x)}dx\right)^{-1}.
\end{equation}

\noindent One other useful relation for the strain distribution follows from \eqref{eq:tensForceBalSI}
\begin{equation}\label{eq:strainDistrSI}
\epsilon(x)=\epsilon_\t{avg}/v(x).
\end{equation}

\noindent For the following calculation, we also introduce a few auxiliary quantities:
\begin{enumerate}
	\item Mean beam width
	\begin{equation}
		w_0=\frac{1}{l}\int_0^l w(x)dx
	\end{equation}
	\item Relative width variation function
	\begin{equation}
		v(x)=w(x)/w_0
	\end{equation}
	\item Static stress $\sigma_{\mathrm{avg}}$ and strain $\epsilon_{\mathrm{avg}}$, averaged over the beam volume
	\begin{align}
		&\sigma_{\mathrm{avg}}=E\,\epsilon_{\mathrm{avg}},\\
		&\epsilon_{\mathrm{avg}}=\frac{1}{h w_0l}\int_0^l h w(x)\epsilon(x)dx= \frac{T}{w_0h}.
	\end{align}
\end{enumerate}\par

\subsection{Vibrational modes}
In order to proceed with explicit calculation the dissipation dilution factors, we first need to find the eigenfrequencies $\omega_n$ and the vibrational mode shapes $u_n$ of a beam. For an elastic beam with high aspect ratio ($l/h$ and $l/w$ much larger than one), these quantities can be found by solving the Euler-Bernoulli equation\cite{landau_theory_1970_SI}
\begin{equation}\label{eq:EBphysSI}
	\frac{d^2}{dx^2}\left(I(x)E\frac{d^2 u_n}{dx^2}\right)-T\frac{d^2 u_n}{dx^2}-\rho_l(x)\omega_n^2 u_n=0,
\end{equation}
\noindent where $n$ is the mode index, $\rho_l(x)=\rho h w(x)$ is the linear mass density and $I(x)$ is the geometric moment of inertia. In order to simplify the notation, it is convenient to introduce a normalized length, $s=x/l$, taking values from 0 to 1, and use it to transform \eqref{eq:EBphysSI} to a new form
\begin{equation}\label{eq:EBSI}
	\lambda^2\frac{1}{v(s)}\frac{d^2}{ds^2}\left(v(s)\frac{d^2 u_n}{ds^2}\right)-\frac{1}{v(s)}\frac{d^2 u_n}{ds^2}-\Omega_n^2 u_n=0,
\end{equation}
\noindent where $\Omega$ is the dimensionless frequency
\begin{equation}\label{eq:dimlessFreqDefSI}
\Omega^2=\frac{\rho l^2 \omega^2}{\epsilon_{\t{avg}}E},
\end{equation}
\noindent and $\lambda$ is the strain dilution parameter given by
\begin{equation}\label{eq:lambdaDefSI}
\lambda^2=\frac{1}{12\epsilon_{\t{avg}}}\frac{h^2}{l^2}.
\end{equation}
The high-strain limit corresponds to $\lambda$ being much smaller than 1. For a doubly clamped beam, the eigenvalue problem in \eqref{eq:EBSI} is supplemented with boundary conditions
\begin{equation}
	u(0)=u(1)=0,\, u'(0)=u'(1)=0.
\end{equation}
	
\subsection{Derivation of distributed and clamping losses}
The evaluation of the integrals in \eqref{eq:WelTensSI}-\ref{eq:WelBendSI} provides us with a general formula for the dissipation dilution of the $n$-th mode
\begin{equation}\label{eq:DQIntFormSI}
	D_{Q,n}=1+\frac{1}{\lambda^2}\frac{\int_{0}^{1}\left(u_n'(s)\right)^2ds}{\int_{0}^{1}v(s)\left(u_n''(s)\right)^2ds},
\end{equation}
in which one can separate the contributions due to the distributed and the clamping parts of the mode. Here we are interested in the strong dilution limit, where $D_Q\gg 1$ and therefore we neglect the first term in \eqref{eq:DQIntFormSI}. In order to find the distributed energy, we neglect the bending term in \eqref{eq:EBSI}, which only weakly perturbs the solution in the region away from the clamping points (see \figref{fig:bendIllSI}A), and find the mode shapes $u_n$ from
\begin{equation}\label{eq:waveSI}
	-\frac{1}{v(s)}\frac{d^2 u_n(s)}{ds^2}=\Omega^2_n u_n(s).
\end{equation}
The tension and bending energy integrals in \eqref{eq:DQIntFormSI} can be transformed to a new form
\begin{align}
	&\int_{0}^{1}\left(u_n'(s)\right)^2ds=\Omega^2_n\int_{0}^{1}v(s)u_n(s)^2ds,\\
	&\int_{0}^{1}v(s)\left(u_n''(s)\right)^2ds=\Omega^4_n\int_{0}^{1}v(s)^3 u_n(s)^2ds.\label{eq:bendingDistributed}
\end{align}
In addition to the distributed contributions given by \eqref{eq:bendingDistributed}, the bending energy includes a contribution from the clamping regions. The tensile energy stored in these regions is negligibly small. Near the clamping points the bending term in Euler-Bernoulli equation is significant due to the boundary condition $u'(0)=u'(1)=0$, but, on the other hand, $u$ is close to $0$ so that $\Omega_n^2 u$ can be neglected. In the region around $s=0$, assuming that the beam width is approximately constant here, such that $v(s)=v_{\mathrm{cl}}$, \eqref{eq:EBSI} reduces to
\begin{equation}
	\lambda^2 v_{\mathrm{cl}}u''''(s)-u''(s)=0.
\end{equation}
The general solution is given by
\begin{equation}
	u(s)=C_1+C_2 s+ C_3 e^{-s/(\lambda\sqrt{ v_{\mathrm{cl}}})}+ C_4 e^{s/(\lambda\sqrt{ v_{\mathrm{cl}}})},
\end{equation}
where the constants $C_{1-4}$ can be found from the boundary conditions: $u(0)=0$, $u'(0)=0$ and $u'(s\gg \lambda\sqrt{ v_{\mathrm{cl}}})=u_{\t{cl},n}'$. For the solution, $u_n$, to the wave equation given by ~\eqref{eq:waveSI}, $u_\t{cl,n}'=u_n'(0)$. $u_n$ does therefore not satisfy the boundary condition $u_n'(0)=0$ per se. Explicitly,
\begin{equation}\label{eq:clampSolSI}
	u(s)=u_{\t{cl},n}'\left(s+ \lambda\sqrt{v_{\mathrm{cl}}}\left(e^{-s/(\lambda\sqrt{ v_{\mathrm{cl}}})}-1\right)\right).
\end{equation}
and the contribution of the clamping point into the curvature integral is found as
\begin{equation}\label{eq:clampLossSI}
	\int_{0}^{\infty}v(s)\left(u''(s)\right)^2ds=\frac{1}{2\lambda}\sqrt{v_{\mathrm{cl}}}(u_{\t{cl},n}')^2.
\end{equation}
Note, that the clamping region is small $\Delta x_\t{cl}/l=\lambda\sqrt{v_{\mathrm{cl}}}\ll 1$ and the bending energy stored here is proportional to the magnitude of the mode envelope at the beam boundaries. Combining the clamping (assumed to be equal at both clamping points, $s=0$ and $s=1$) and the distributed contributions, we arrive at
\begin{equation}\label{eq:DQgeneralSI}
	D_{Q,n}=\frac{1}{2\alpha_n\lambda+\beta_n\Omega^2_n\lambda^2},
\end{equation}
where
\begin{align}
&\alpha_n=\frac{\sqrt{v_{\mathrm{cl}}}(u_{\t{cl},n}')^2}{2\Omega^2_n\left(\int_0^1v(s)u_n(s)^2ds\right)},\label{eq:alphaDefSI}\\
&\beta_n=\frac{\int_{0}^{1}v(s)^3 u_n(s)^2ds}{\int_0^1v(s)u_n(s)^2ds}.\label{eq:betaDefSI}
\end{align}

With the help of \eqref{eq:DQgeneralSI}-\ref{eq:betaDefSI}, the optimization of dissipation dilution can be performed by shaping $v(s)$ to reduce $\alpha_n$ (clamping losses) and $\beta_n$ (distributed losses). For a uniform rectangular beam $v(s)=1$, $\Omega^2_n=(\pi n)^2$ and $u_n=\sqrt{2}\sin(\pi n\,s)$, which yields $\alpha_n=1$, $\beta_n=1$ and reproduces the result from~\cite{gonzalez_brownian_1994_SI}
\begin{equation}\label{eq:DQRectBeamSI}
	D_{Q,n}^\t{rect.beam}=\frac{1}{2\lambda+(n\pi)^2 \lambda^2}.
\end{equation}

\begin{figure*}[t]
	\centering
	\includegraphics[width=\textwidth]{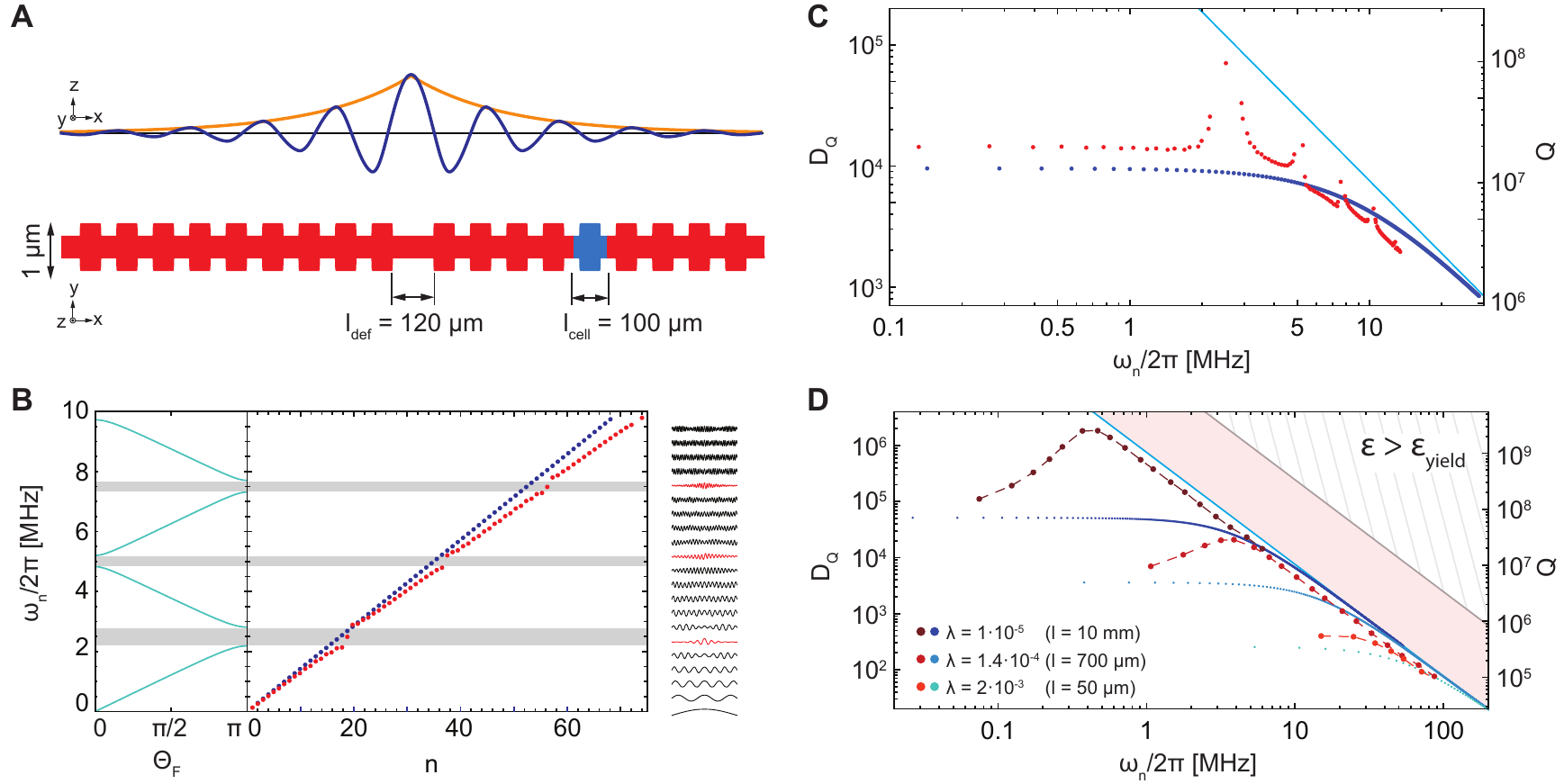}
	\caption{A) Top: localized flexural mode shape $u(x)$ (blue) and its exponential envelope (orange). Bottom:  Geometry of a beam (red). PnC unit cell is highlighted blue within the beam. B) Red dots: spectrum of the out-of-plane flexural vibrations of the beam shown in panel A. Blue lines: band diagram of a phononic crystal with the cell highlighted in panel A. Blue dots: spectrum of a uniform rectangular beam with same $l,\,h$ as the PnC beam. C) Dissipation dilution of PnC beam modes (red dots) compared to the modes of a uniform beam (blue dots) and a uniform beam without clamping losses (blue line). D) Variation of localized mode $D_Q$ (shades of red) with frequency and beam length, and comparison to modes of uniform beams of the same lengths. Beam thickness here is $h=20$ nm. Localized mode frequency is changed by the variation of the number of PnC unit cells within the beam (together with the unit cell length as the beam length is fixed) while keeping the ratio of the central defect to the unit cell length constant.}
	\label{fig:softClampSI}
\end{figure*}

\section{Absolute quality factors of Si$_3$N$_4$ nanobeams}
If the resonator dissipation is due to intrinsic losses, the absolute mode quality factors can be calculated according to \eqref{eq:QIntDilSI} from the intrinsic material quality factor $Q_\t{int}$ and $D_{Q}$ as
\begin{equation}\label{eq:QabsSI}
Q=D_{Q}\times Q_\t{int}.
\end{equation}
In the high-strain limit ($\lambda\ll 1$) $D_{Q}$ depends only on the beam geometry, mode order and strain, but not on any of the material parameters. Dissipation dilution can therefore be understood without ever specifying a material. However, we present calculations of absolute $Q$ factors assuming the material is stoichiometric Si$_3$N$_4$, as it is by far the most popular platform for strained high-$Q$ nanomechnical resonators (see, for example Villanueva et al.\cite{villanueva_surface_loss_2014_SI}, and references therein). In particular, we assume parameters consistent with the Si$_3$N$_4$ deposited by low pressure chemical vapor deposition, as used in\cite{ghadimi_strain_2017_SI}: deposition strain $\epsilon_{\t{film}}=0.46\%$ (stress $\sigma_\t{film}=1.14$ GPa), Young's modulus $E=250$ GPa, Poisson's ratio $\nu=0.23$ and density $\rho=3100$ kg/m$^3$.\par

The intrinsic quality factor of Si$_3$N$_4$ was found to be almost frequency independent within the range $100$ kHz--$50$ MHz\cite{fedorov_evidence_2018_SI} but increasing with thickness due to surface losses. For Si$_3$N$_4$ of smaller thickness than 100 nm it was phenomenologically established\cite{villanueva_surface_loss_2014_SI,tsaturyan_ultracoherent_2017_SI,ghadimi_strain_2017_SI} that the intrinsic quality factor is proportional to thickness.
\begin{equation}
	Q_\t{int}(h)=6900\frac{h}{\mathrm{[100\,nm]}}.
\end{equation}
	
\section{Soft clamped modes in phononic crystal beams}\label{sec:dissDilEngSI}

\begin{figure}[t]
	\centering
	\includegraphics[width=0.5\textwidth]{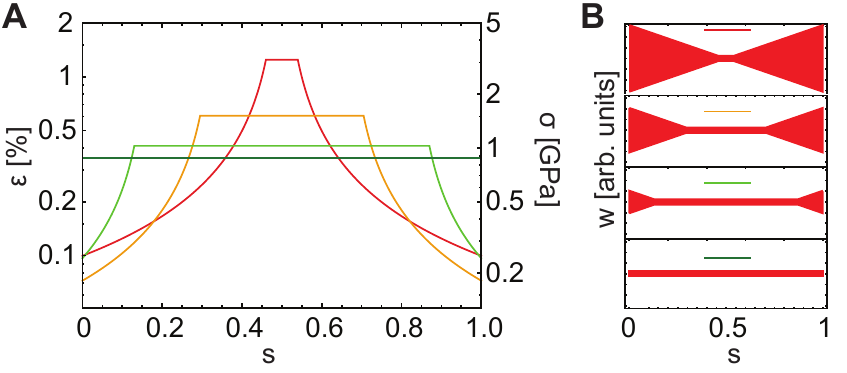}
	\caption{A) Distribution of axial strain and stress along beams with strain-enhancing fillets, with geometries corresponding to the panel B). Here $l_\t{waist}/l_\t{tot}=0.08,\, 0.41,\, 0.74,\, 1$ for red, orange, light- and dark green curves correspondingly. These geometries feature reduced dissipation dilution as discussed in the main text and only serve for illustration of strain redistribution.}
	\label{fig:strainDistrSI}
\end{figure}

\figref{fig:softClampSI} shows an example of a \SiN PnC nanobeam featuring a soft-clamped vibrational mode for which clamping loss contribution is suppressed. The calculation is made for a 20 nm thick beam that consists of two phononic crystal barriers, each incorporating nine 100 $\mu$m-long unit cells, and a 120 $\mu$m-long central defect region. The band diagram for the unit cell vibrations in the out-of-plane $z$-direction is plotted in \figref{fig:softClampSI}B, showing the frequencies of the stopbands. The mode spectrum of the finite beam from panel A is shown next to the band diagram in \figref{fig:softClampSI}B and it can be seen that a mode localized around the defect exists in the first bandgap. The quality factors of the beam modes, plotted in \figref{fig:softClampSI}C, show improved $Q$ for the localized mode due to the suppression of clamping losses. Here the $Q$s of the first localized mode is approaching the ideal value $Q=Q_\t{int}/(n\pi\lambda)^2$ that a uniform beam mode would have without clamping losses (blue line), while the $Q$s of the modes with frequencies outside of the PnC bandgaps are similar to those of a regular uniform beam (blue points).\par

The relative advantage from using the soft clamping is thus greater for beams with smaller $\lambda$. In order to illustrate this numerically, in \figref{fig:softClampSI}D, we plot the localized mode $D_Q$s and absolute $Q$s as their frequency is varied by changing the number of unit cells within the beam . The center defect length is fixed to be 1.2 of the unit cell length in order to keep the soft clamped mode frequency approximately in the bandgap center. As the dissipation dilution parameter $\lambda$ is varied by changing the beam length from 50 $\mu$m to 10 mm, one can observe that both the maximum absolute $Q$s and the relative $Q$ enhancement of a soft clamped mode compared to uniform beam modes increase dramatically. At the same time, the frequency of the highest-$Q$ localized mode necessarily shifts down with increasing $l$, thus never entering the shaded red area and taking values between
\begin{equation}
\frac{12E\epsilon_{\t{film}}^2}{(1-\nu)^2\rho h^2\omega^2}<D_Q<\frac{12E\epsilon_\t{yield}^2}{\rho h^2\omega^2}.
\end{equation}
The area forbidden by the breaking strain is hatched gray in \figref{fig:softClampSI}D.

\section{Geometrical strain engineering}\label{sec:geomStrainEngSI}

Taking insights from \eqref{eq:strainDistrSI}, it can be seen that shaping the transverse profile of the beam allows the increase of strain in regions where the beam is thinner than on average. To gain some intuition into this geometric strain enhancement, it is useful to consider the evolution of strain distribution in the bowtie structure shown in \figref{fig:strainDistrSI} as the waist is being reduced compare to the triangular supports. It can be seen from the figure that larger supports result in larger peak strain, but smaller spatial extent of the high-strain region in the beam center. This is a general rule that can be made quantitative as follows: strain enhancement in a thin waist between two supports of arbitrary shapes is estimated as
\begin{equation}\label{eq:strainEnhLimSI}
\frac{\epsilon_\t{max}}{\epsilon_\t{avg}}\lesssim \t{min}\left(\frac{l_\t{waist}}{l_\t{tot}}, \frac{w_\t{waist}}{w_\t{supp}}\right).
\end{equation}
\eqref{eq:strainEnhLimSI} is useful not only for the evaluation of geometric strain enhancement in beams, but also in membranes. For example, it predicts that no significant enhancement of strain takes place in the tethers of a trampoline membrane with fillet radius smaller than the tether length.

\begin{figure}[t]
	\centering
	\includegraphics[width=\textwidth]{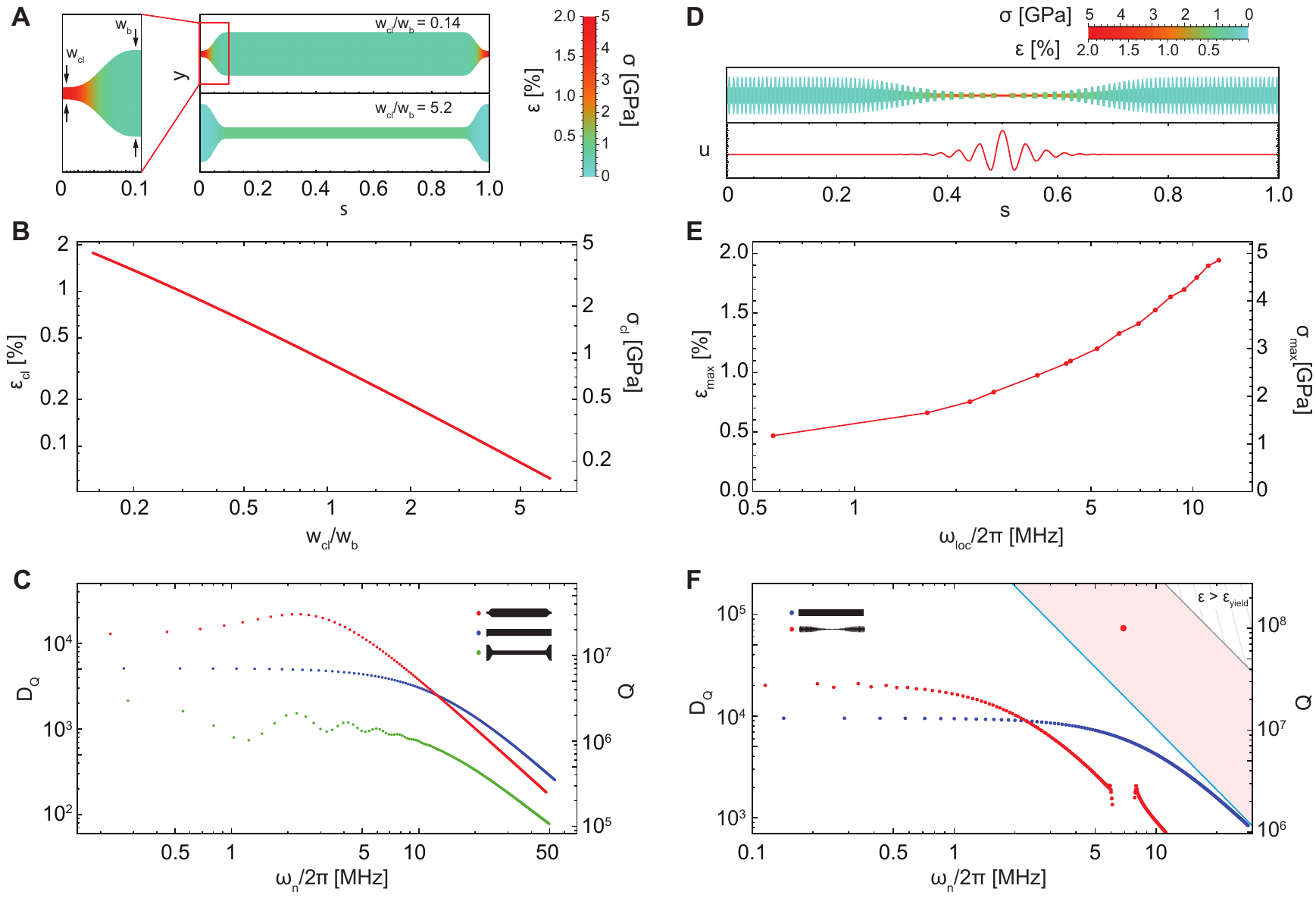}
	\caption{\small{A) Geometry and color-coded strain distribution in beams with thin (top) and thick (bottom) clampings). B) Variation of the strain in the clamping region as a function of clamp width $w_\t{cl}$ relative to the beam center $w_\t{b}$. C) Dissipation dilutions and quality factors of modes of a thin-clamp beam (red), a uniform rectangular beam (blue) and a thick-clamp beam (green). Here $l=1$ mm, $h=20$ nm. D) Top: strain distribution in a non-uniform PnC beam. Bottom: vibrational mode, localized in the high-strain region of the PnC. E) Variation of the maximum strain (in the beam center) with localized mode frequency. F) Red: modes of a $l=3$ mm $h=20$ nm beam with the shape from panel D. Blue: modes of uniform beam with the same $l,\,h$. The egion with red background shows the range of $D_Q$ values that exceed the $D_Q$ of an idealized clampless beam, but is not forbidden by the breaking strain.}}
	\label{fig:scseSI}
\end{figure}

\section{Spectra of beams with local strain enhancement}

Spectra of out-of-plane vibrations of beams with engineered local strain enhancement are shown in \figref{fig:scseSI}. As shown in \figref{fig:scseSI}C, all low-frequency flexural modes of thin-clamping beam have enhanced $Q$s, while the $Q$s of thick-clamping beams are decreased. In \figref{fig:scseSI}C the spectrum of a 3 mm long 20 nm thick tapered PnC beam is shown. As in the case of a bare PnC beam, it features a ``soft clamped" localized mode with dissipation dilution (and thus $Q$) significantly higher than the normal value in a uniform beam, but in addition, here the localized mode $Q$ even exceeds that of an ideal clamp-less beam tensioned by the material strain and enters the area shaded pink. This confirms the increase in effective strain experienced by the soft-clamped mode.\par

\section{Comparison between 1D and 3D simulations of dissipation dilution and non-flexural modes of PnC beams with engineered strain}

In the main text we employ a one dimensional model to treat the case of doubly clamped beams and consider only out-of-plane flexural modes. These modes are our primary interest as they exhibit the highest dissipation dilution factors and, correspondingly, $Q$s. The vibrational spectrum of a beam, however, also includes other mode families like in-plane flexural, torsional and longitudinal modes. To give an example of a complete spectrum, we simulate first 100 modes of a PnC beam with strain engineering using commercial FEM software and calculate the dissipation dilution factors for these modes using \eqref{eq:QIntDilSI}. The results are presented in \figref{fig:MathCOMSOLCmpSI}A. We have to limit the aspect ratio of the geometry in this case to a relatively moderate value set by the length $l=500$ $mu$m and thickness $h=100$ nm as we find that the results of 3D simulation of mode shapes may not be reliable for higher aspect ratios. We also plot in \figref{fig:MathCOMSOLCmpSI}A the predictions of our 1D model for out-of-plane flexural modes, which are in an excellent agreement with the orders-of-magnitude more time consuming full 3D simulations. The most prominent deviation between the 1D and 3D calculations that we observe takes place for a flexural mode presented in \figref{fig:MathCOMSOLCmpSI}D, that happened to hybridize with a low-$Q$ torsional mode and correspondingly have a reduced quality factor.
\begin{figure*}[t]
	\centering
	\includegraphics[width=\textwidth]{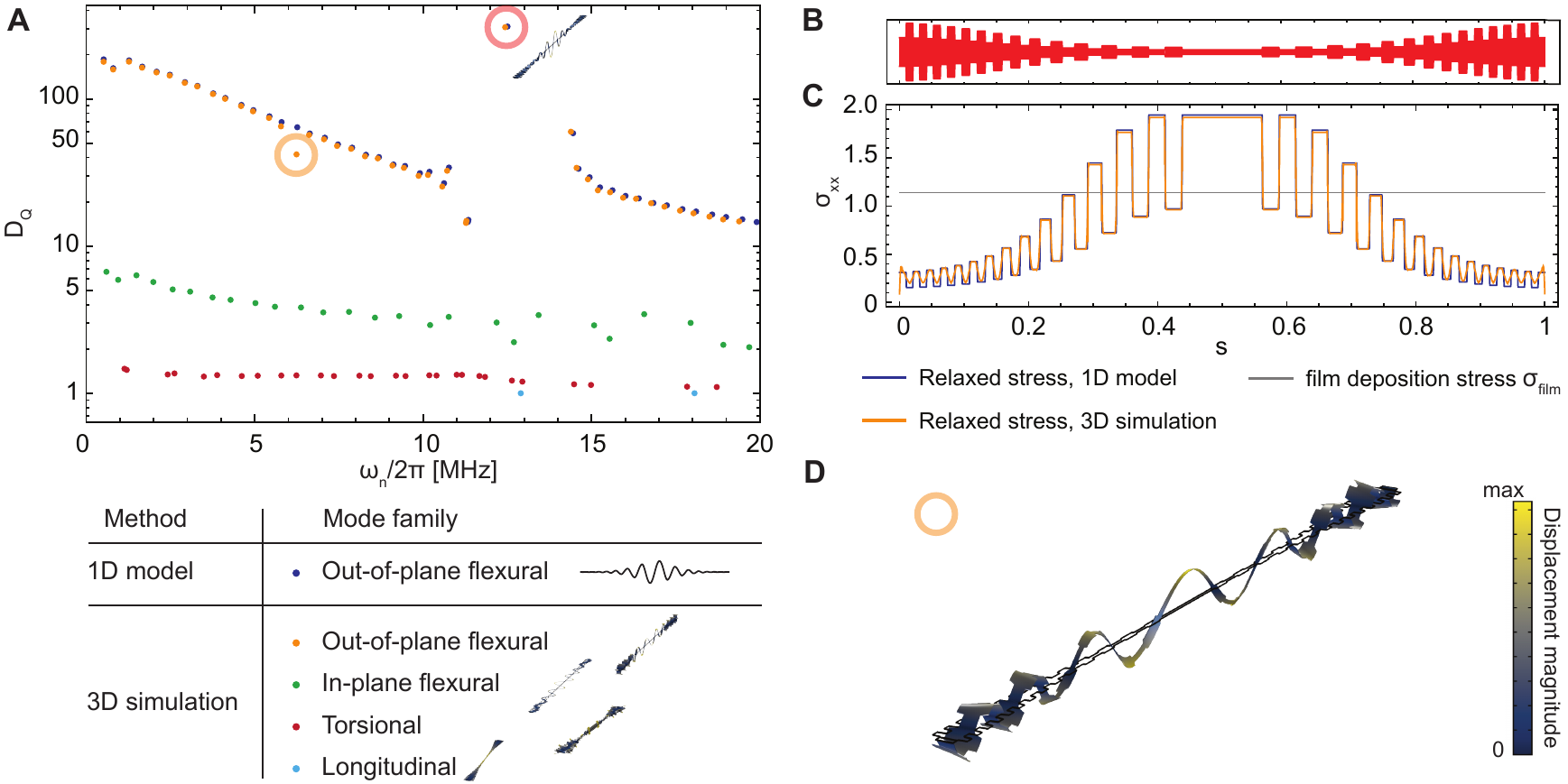}
	\caption{\small{A) Dissipation dilution versus mode frequency calculated using 3D vibrational mode shapes for the first 100 modes of a doubly-clamped beam with length $l=500$ $\mu$m, thickness $h=100$ nm, center width $w=400$ nm and transverse profile shown in B). Orange, green, red and blue dots correspond to different mode families as explained in the caption. For comparison, calculations using the 1D model \cite{mathematica_package_SI} are also presented for out-of-plane flexural modes (dark blue dots). The red circle highlights the localized ``soft clamped" mode. The orange circle highlights a hybridized flexural-rotational mode, the shape of which is shown in D). C) Equilibrium axial stress along the beam center as calculated with the 1D model\cite{mathematica_package_SI} and 3D simulation software.}}
	\label{fig:MathCOMSOLCmpSI}
\end{figure*}


\begin{thebibliography}{40}
\bibitem [{mat()}]{mathematica_package}%
  \BibitemOpen
  \href@noop {} {}\bibinfo {howpublished} {Mathematica package is available at
  zenodo.com, DOI:10.5281/zenodo.1296925}\BibitemShut {NoStop}%
\end{thebibliography}
\end{document}